\documentclass{aa}  

\usepackage{natbib}	
\usepackage{graphicx}
\usepackage{txfonts,epsfig,graphicx,url,twoopt}
\usepackage[breaklinks=true]{hyperref} 
\hypersetup{colorlinks=true,citecolor=blue,linkcolor=blue,urlcolor=red}
\bibpunct{(}{)}{;}{a}{}{,}     
\usepackage[usenames,dvipsnames]{color}

\definecolor{dgreen}{rgb}{0.0,0.5,0}

\usepackage{xcolor}

\begin{document}
   \title{Can dust coagulation trigger streaming instability?}

   \author{J. Dr\k{a}\.{z}kowska\thanks{Member of the International Max Planck Research School for Astronomy and Cosmic Physics at the Heidelberg University}
          \and
          C.P. Dullemond
          }

   \institute{Heidelberg University, Center for Astronomy, Institute of Theoretical Astrophysics, Albert-Ueberle-Str.~2, 69120 Heidelberg, Germany\\
         \email{drazkowska@uni-heidelberg.de}
             }

   \date{Received 14 August 2014 \slash~Accepted 7 October 2014}


 
  \abstract
   {
   Streaming instability can be a very efficient way of overcoming growth and drift barriers to planetesimal formation. However, it was shown that strong clumping, which leads to planetesimal formation, requires a considerable number of large grains. State-of-the-art streaming instability models do not take into account realistic size distributions resulting from the collisional evolution of dust.
   }
   {
    We investigate whether a sufficient quantity of large aggregates can be produced by sticking and what the interplay of dust coagulation and planetesimal formation is.
   }
   {
   We develop a semi-analytical prescription of planetesimal formation by streaming instability and implement it in our dust coagulation code based on the Monte Carlo algorithm with the representative particles approach.
   } 
   {
   We find that planetesimal formation by streaming instability may preferentially work outside the snow line, where sticky icy aggregates are present. The efficiency of the process depends strongly on local dust abundance and radial pressure gradient, and requires a super-solar metallicity. If planetesimal formation is possible, the dust coagulation and settling typically need $\sim$100 orbits to produce sufficiently large and settled grains and planetesimal formation lasts another $\sim$1000 orbits. We present a simple analytical model that computes the amount of dust that can be turned into planetesimals given the parameters of the disk model.
   }
   {}


   \maketitle

\section{Introduction}

The number of known exoplanets is steadily rising, and it already exceeds 1800\footnote{1832 known exoplanets according to the Extrasolar Planets Encyclopaedia \url{http://exoplanet.eu} on 13 October 2014}. Indirect methods indicate that every star in our Galaxy has at least one planet on average \citep{2012Natur.481..167C}. Planet formation is ubiquitous and results in a variety of planetary system architectures. However, the details remain a mystery, as they are impossible to follow with direct observations, which only cover the very first stages of planet formation, when protoplanetary disks consist mainly of gas and small dust particles, and the result in the form of the exoplanet population.

Planets form in disks surrounding young stars. The solid material initially consists only of $\mu$m-sized grains, which are already present in the interstellar medium \citep{2010Sci...329.1622P}. 
Analytical and numerical models of dust evolution aim to explain the growth of the primordial $\mu$m-sized grains through around 40 orders of magnitude in mass to planets that are greater than 1000~km in size. However, the growth already encounters serious obstacles at the beginning of the size range, and these are referred to as growth barriers. There are, in principle, two kinds of growth barriers: those resulting from collisional physics of dust aggregates, and those resulting from radial drift timescale. The first type are known as the bouncing and fragmentation barriers. They arise when the impact speeds are too high to allow aggregate sticking. An overview of collisional physics of dust aggregates is given by \citet{2010A&A...513A..56G}. Implementing a complex model derived from laboratory work into a dust coagulation code, \citet{2010A&A...513A..57Z} found that silicate grain growth is inhibited by aggregate compaction and bouncing already at millimeter sizes. Icy particles, which can exist outside the snow line, are considered to be more sticky and presumably avoid the bouncing behavior \citep{2011ApJ...737...36W}. However, even without bouncing, impact velocities reaching a few tens m~s$^{-1}$ are too high to allow grain growth and instead lead to fragmentation \citep{1993Icar..106..151B, 2008ARA&A..46...21B, 2013MNRAS.435.2371M}. At even smaller sizes, for $<$mm-sized aggregates, there was another sticking barrier found, called the charge barrier, which results from the electrostatic charge of small aggregates \citep{2009ApJ...698.1122O}.

Barriers of the second kind are connected to the radial drift, which is caused by the sub-Keplerian rotation of pressure supported gas disk. The dust grains interact with the gas via aerodynamic drag, lose their angular momentum, and drift toward the central star. For roughly decimeter-sized grains, the drift timescale is typically shorter than the growth timescale. Thus, such grains are removed from a given location in the disk before they can produce any larger aggregates \citep{1977MNRAS.180...57W, 1986Icar...67..375N, 2007A&A...469.1169B, 2008A&A...480..859B}. 

There are several concepts that facilitate overcoming the growth barriers and producing bodies which are held together by self-gravity, for example planetesimals with minimum size of 100~m \citep{1999Icar..142....5B}. Such concepts are, among others, pressure bumps \citep{1972fpp..conf..211W, 2007ApJ...664L..55K, 2008A&A...487L...1B, 2012A&A...538A.114P, 2013A&A...556A..37D}, dust accumulation in vortices \citep{1995A&A...295L...1B, 1999PhFl...11.2280B, 2006ApJ...639..432K, 2009A&A...497..869L, 2010A&A...516A..31M}, sweep-up growth scenario \citep{2012A&A...540A..73W, 2012A&A...544L..16W, 2013ApJ...764..146G, 2013ApJ...774L...4M, 2014A&A...567A..38D}, ice condensation \citep{2004ApJ...614..490C,2013A&A...552A.137R}, and ultra-porous grain growth \citep{2012ApJ...752..106O, 2013A&A...557L...4K}. A review of the current understanding of planetesimal formation can be found in \citet{2014arXiv1402.1344J}. In this paper, we focus on planetesimal formation by gravitational collapse of dense dust clumps produced by two-fluid instability known as the streaming instability.

One of the early scenarios of planetesimal formation was proposed by \citet{1973ApJ...183.1051G}. This model relied on dust settling toward the midplane of the protoplanetary disk and the formation of a~thin, unstable disk of solids. This thin disk would then fragment because of the gravitational instability and the fragments would collapse directly into planetesimals. However, it was found that formation of such a thin midplane layer is not possible, because as soon as the local dust-to-gas ratio in the midplane exceeds unity, the shear instabilities occur \citep{1980Icar...44..172W,1993Icar..106..102C,1995Icar..116..433W}, in particular the Kelvin-Helmholtz instability \citep{2006ApJ...643.1219J,2009ApJ...691..907B}. 

Formation of planetesimals by gravitational instability is still possible if a strong clumping of dust is present. The complicated interactions between gas and dust may lead to the development of a powerful instability, called streaming instability, that causes strong inhomogeneities in the dust density \citep{2000Icar..148..537G,2005ApJ...620..459Y}. The instability is most efficient for high dust-to-gas ratios and large particles, with stopping times corresponding to the orbital timescale. \citet{2007ApJ...662..613Y} and \citet{2007Natur.448.1022J} confirmed that this instability is able to produce very dense dust clumps and leads to rapid planetesimal formation. This result was later confirmed by high resolution numerical simulations \citep{2011A&A...529A..62J}, as well as by other authors using different methods and codes \citep{2009MNRAS.397...24B,2010MNRAS.403..211T,2010ApJ...722.1437B,2011MNRAS.415.3591J,2013MNRAS.434.1460K}.

The numerical models of the streaming instability are generally computationally expensive and do not allow us to perform wide parameter studies or use extended simulation domains. They are also typically initialized with an arbitrary particle size distribution, often with even just a single particle species. The aggregates sizes used in the simulations are usually large, corresponding to Stokes numbers of $10^{-2}-1$, whereas in a primordial protoplanetary nebula we expect $\mu$m-sized grains, which correspond to Stokes numbers of $10^{-6}$. Larger grains may be produced by coagulation, but their maximum size is limited by the growth barriers at Stokes numbers in the range $10^{-4}-10^{-1}$. Choosing only a narrow size distribution and assuming that all of the dust particles are large leads to a kind of super-critical initial condition and the instability is triggered very quickly.
We expect that gradual dust growth leads to a quite wide and highly problem-dependent size distribution, meaning that only a fraction of the largest grains will be able to participate in clumping \citep{2010ApJ...722.1437B}. These particles can already form planetesimals, while other particles are still growing and gradually refill the population of big grains. 

In addition to the abundant large grains, streaming instability also requires that the vertically integrated dust-to-gas ratio is super-solar and the local dust-to-gas ratio is higher than unity, which means that it can only happen in a dense midplane layer. Such a midplane layer can be relatively easily formed in a dead zone, where no turbulent diffusion is present. 
In order to investigate the interplay of dust coagulation and planetesimal formation in the streaming instability triggered in a dead zone of protoplanetary disks, we build a semi-analytical model of the latter effect into our dust coagulation code based on the Monte Carlo algorithm and the representative particle approach \citep{2008A&A...489..931Z, 2013A&A...556A..37D}. We base our model of the streaming instability on the work of \citet{2010ApJ...722.1437B, 2010ApJ...722L.220B}. We describe the numerical model in Sect.~\ref{sub:model}. Basing our work on this approach, we present simple estimates in Sect.~\ref{sub:pre} and results of the full numerical models in Sect.~\ref{sub:results}. We offer an analytical formula that explains these results in Sect.~\ref{sub:expl}. We discuss the limitations of our approach in Sect.~\ref{sub:discuss} and summarize our work in Sect.~\ref{sub:last}.  

\section{Numerical methods}\label{sub:model}

\subsection{Dust evolution}

We use the Monte Carlo method and the representative particle approach first described by \citet{2008A&A...489..931Z} to model dust evolution. 
We developed this method by adding an adaptive grid routine, which enables multi-dimensional models. The code was presented by \citet{2013A&A...556A..37D}. For this paper, we focus on local models, only including the vertical dimension. We have already tested our code with such setups \cite[see Sect.~4 of][]{2013A&A...556A..37D} and we found a good agreement with the results of \citet{2005A&A...434..971D} and \citet{2011A&A...534A..73Z}. We have also noticed that the adaptive grid routine allows us to obtain very high resolution of the midplane layer and thus to follow dust evolution in dead zones well. 

The representative particle approach relies on the assumption that the evolution of many dust particles can be resolved by following the evolution of only a limited number of so called representative particles. One representative particle represents a swarm of identical physical particles. In our code, all the swarms have equal mass, thus the mass of one swarm $M_{\rm{swarm}}$ is an adequate fraction of the total dust mass $M_{\rm{tot}}$
\begin{equation}\label{mswarm}
M_{\rm{swarm}} = M_{\rm{tot}} / N_{\rm{swarms}},
\end{equation} 
where $N_{\rm{swarms}}$ is the number of representative particles, or swarms, present in our simulation \citep[for discussion of the approach restrictions, see][]{2014A&A...567A..38D}. The collisions between dust particles are modeled with a Monte Carlo algorithm, described in detail in \citet{2008A&A...489..931Z} and \citet{2013A&A...556A..37D}. 

In the original code, the particles were influenced by the gas but the gas did not change its state. Here, we assume that the particles are active, i.e., we enable the back-reaction on gas. The effects of the back-reaction are implemented in a semi-analytical way. 
First, a very crude approach of implementing the turbulence triggered by the streaming instability was already made in the \citet{2013A&A...556A..37D}. However, that approach did not include the dust clumping and the possibility of planetesimal formation, which we include now. As we do not directly model the hydrodynamics of the gas, we cannot capture the two-fluid interactions explicitly. We model dust growth and settling, and include the effects of back-reaction based on results of the direct numerical simulations that investigated the dynamics of dust grains in the midplane of protoplanetary disks presented by \citet{2010ApJ...722.1437B,2010ApJ...722L.220B}, and the analytical model by \citet{2012ApJ...744..101T}. We describe this approach in Sect.~\ref{sub:si}. 

The dust evolution in a protoplanetary disk is driven by its interactions with gas. We assume that the gas disk is described by the minimum mass solar nebula (henceforth MMSN) model proposed by \citet{1981PThPS..70...35H}, where the surface density follows
\begin{equation}
\Sigma_{\rm g}=1700\times\left(\frac{r}{\rm AU}\right)^{-1.5} {\rm g~cm}^{-2},
\end{equation}
where $r$ is radial distance to the central star of mass $1~M_\odot$. The surface density of dust is parametrized by metallicity\footnote{In this paper, we use the term {\it metallicity} interchangeably with the {\it vertically integrated dust-to-gas ratio}.} $Z$ such that $\Sigma_{\rm d}=Z\times\Sigma_{\rm g}$.
The vertical structure of the gas is described by the local hydrostatic equilibrium and thus the gas density follows
\begin{equation}
\rho_{\rm g}(z)=\frac{\Sigma_{\rm g}}{\sqrt{2\pi}H_{\rm g}}\exp\left(\frac{-z^2}{2H_{\rm g}^2}\right),
\end{equation}
where $z$ is the distance to the midplane and $H_{\rm g}=c_{\rm s}{\slash}\Omega_{\rm K}$ is the pressure scale height of gas defined by the sound speed $c_{\rm s}$ and orbital frequency $\Omega_{\rm K}$. We assume an isothermal disk with temperature
\begin{equation}
T = 280\times\left(\frac{r}{\rm AU}\right)^{-0.5} {\rm K}.
\end{equation}

To investigate dust particles dynamics, it is convenient to use the Stokes number defined as
\begin{equation}
{\rm St}=t_{\rm{s}} \Omega_{\rm{K}},
\end{equation}
where $t_{\rm{s}}$ is the stopping time of the dust particle. The stopping time of a particle determines the timescale that the particle needs to adjust its velocity to the velocity of the surrounding gas. The exact expression that we use to compute the $t_{\rm{s}}$ depends on the particle radius, and we use the formulas given by \citet{1977MNRAS.180...57W}.
The Stokes number can be used as a particle-gas coupling strength indicator. The particles with $\rm{St} \ll 1$ are completely coupled and follow the motion of the gas, whereas the particles with $\rm{St} \gg 1$ are fully decoupled from the gas. The particles that have $\rm{St} \approx 1$, sometimes called pebbles, are marginally coupled to the gas and suffer the most from this interaction by acquiring high drift and impact speeds. However, these are also the grains that can trigger the streaming instability and form planetesimals, which is what we are investigating in this paper.

In a laminar disk, dust settles because of the vertical component of the star gravity and gas drag. We use the vertical velocity that implements the damped oscillations of large grains, such that the sedimentation rate is consistent with the results of \citet{2011MNRAS.415...93C}:
\begin{equation}\label{vsett}
v_{\rm{z}} = -z \Omega_{\rm{K}} \frac{\rm St}{1+{\rm St}^2}.
\end{equation}
If turbulence is present, we use the $\alpha$ prescription \citep{1973A&A....24..337S}, and the turbulent mixing of dust is implemented as random kicks \citep{2010ApJ...723..514C}.
The height of the dust layer is regulated by the turbulence strength and settling. Including orbital oscillations of grains with ${\rm St}>1$, we get the scale height of the dust layer \citep{2006MNRAS.373.1633C, 2007Icar..192..588Y, 2011MNRAS.415...93C}
\begin{equation}\label{hdust}
H_{\rm{d}} = H_{\rm{g}} \sqrt{\frac{\alpha}{\alpha+\rm{St}}}.
\end{equation}

The collisions between dust aggregates are driven by the Brownian motion, relative radial, transverse and vertical drift, and turbulence. Although we do not include the radial drift of particles explicitly, we keep the drift speeds as a source of impact velocities, to correctly capture the growth rates and collisional physics.

\subsection{Streaming instability}\label{sub:si}

In a laminar disk, all dust grains would settle to the midplane and form a very thin and gravitationally unstable midplane layer \citep{1981Icar...45..517N}. However, the shear between such a layer and gas triggers turbulence that maintains its height at a level that typically does not allow direct gravitational collapse. 
The strength of turbulence triggered by the midplane instability was recently investigated by \citet{2012ApJ...744..101T}. They estimated the turbulence efficiency from the energy supplied by the radial drift of dust. Irrespectively of the turbulence mechanism, which can be both the Kelvin-Helmholz and streaming instability, they find that for particles with ${\rm St}<1$ the resulting turbulence strength can be parametrized with
\begin{equation}\label{alpha}
\alpha = \left[\left({\rm C}_1{\rm C}_{\rm eff}\eta Z\right)^{-2/3}+\left({\rm C}_2 {\rm C}_{\rm eff}\eta Z^{-1}\right)^{-2}\right]^{-1}{\rm St},
\end{equation}
where ${\rm C}_1=1$, ${\rm C}_2=1.6$, and ${\rm C_{\rm eff}}=0.19$. As the derivation of this equation was based on the assumption that all dust grains have an equal size corresponding to one Stokes number ${\rm St}$, in our implementation we take ${\rm St = \bar{St}}$, a mass-weighted average Stokes number of all the particles. Parameter $\eta$ is related to the gas midplane pressure $P_{\rm g}$ gradient
\begin{equation}\label{eta}
\eta = \frac{1}{2\rho_{\rm g}r\Omega_{\rm K}^2} \frac{\partial P_{\rm g}}{\partial r}.
\end{equation}
It is useful to parametrize the pressure gradient as
\begin{equation}\label{pi}
\Pi = \frac{|\eta| v_{\rm K}}{c_{\rm s}},
\end{equation}
the ratio of maximum radial drift speed $|\eta| v_{\rm K}$, with $v_{\rm K}$ being the orbital velocity, and the isothermal sound speed $c_{\rm s}$.

\citet{2010ApJ...722.1437B,2010ApJ...722L.220B} (henceforth BS10) performed local 2-D and 3-D numerical simulations of particles and gas dynamics in the midplane of a protoplanetary disk using the Athena code, including gas as a fluid and dust as superparticles \citep{2010ApJS..190..297B}. Both the aerodynamic coupling and particles to gas feedback were included. Magnetic forces and self-gravity were ignored. They focused on a laminar disk, where the turbulence is initially not present. 
They regulated the aggregate sizes by varying the minimum and maximum Stokes number of particles and by assuming that each logarithmic particle size bin is represented by the same amount of mass.
They found that the settling of particles triggers the streaming instability and the related turbulence maintained the particles height before the Kelvin-Helmholtz instability could emerge. The strength of this turbulence is comparable with the results of \citet{2012ApJ...744..101T} represented by Eq.~(\ref{alpha}).

The most interesting property of the streaming instability in the planet formation context is its ability to concentrate dust particles in dense clumps. \citet{2010ApJ...722.1437B, 2010ApJ...722L.220B} performed a parameter study, varying dust sizes, the dust-to-gas ratio, and the pressure gradient, in order to define threshold conditions for the strong clumping, over the Roche density, which could lead to planetesimal formation. The grain size distributions used by BS10 had the minimum Stokes number of 10$^{-4}$ and the maximum of 1. Analyzing the results of their runs, in particular the dense clumps composition, they concluded that the smallest particles needed to trigger such a strong clumping correspond to a critical Stokes number of ${\rm St_{crit}}=10^{-2}$ and that a super-solar dust abundance is required. They found that a higher vertically integrated dust-to-gas ratio $Z_{\rm tot}$ lowers the threshold abundance of large grains, as a high dust-to-gas ratio reduces the turbulence and makes it easier to form a dense midplane layer. They also found that the lower the radial pressure gradient, the more easily the streaming instability can be triggered, consistent with findings of \citet{2007Natur.448.1022J}.  

The BS10 work suggests a critical total metallicity as a criterion for strong clumping and they find its value to be in the range $0.02-0.07$ for different values of the pressure gradient and different sizes of dust grains. They assumed a flat mass distribution in logarithmic size bins in their models. In our runs, the size distribution is an outcome of the Monte Carlo dust coagulation modeling. In order to use the BS10 results in our code, we build a model relying on splitting the dust mass distribution in two parts: particles larger and smaller than the size corresponding to the critical value of the Stokes number ${\rm St_{crit}}=10^{-2}$. In our model, we calculate the metallicity $Z({\rm St}>10^{-2})$ taking into account only the large grains, with Stokes number higher than ${\rm St_{crit}}=10^{-2}$,
\begin{equation}\label{zcritdef}
Z({\rm St}>10^{-2}) = \sum_{\rm St>10^{-2}}\frac{\Sigma_{\rm d}({\rm St})}{\Sigma_{\rm g}},
\end{equation}
where $\Sigma_{\rm d}({\rm St})$ is dust surface density contributed by particles corresponding to a Stokes number ${\rm St}$.
If the abundance of large grains $Z({\rm St}>10^{-2})$ is higher than a critical value $Z_{\rm crit}$, we assume that clumping over the Roche density happens and planetesimal formation is possible.

\begin{figure}
   \centering
   \includegraphics[width=0.95\hsize]{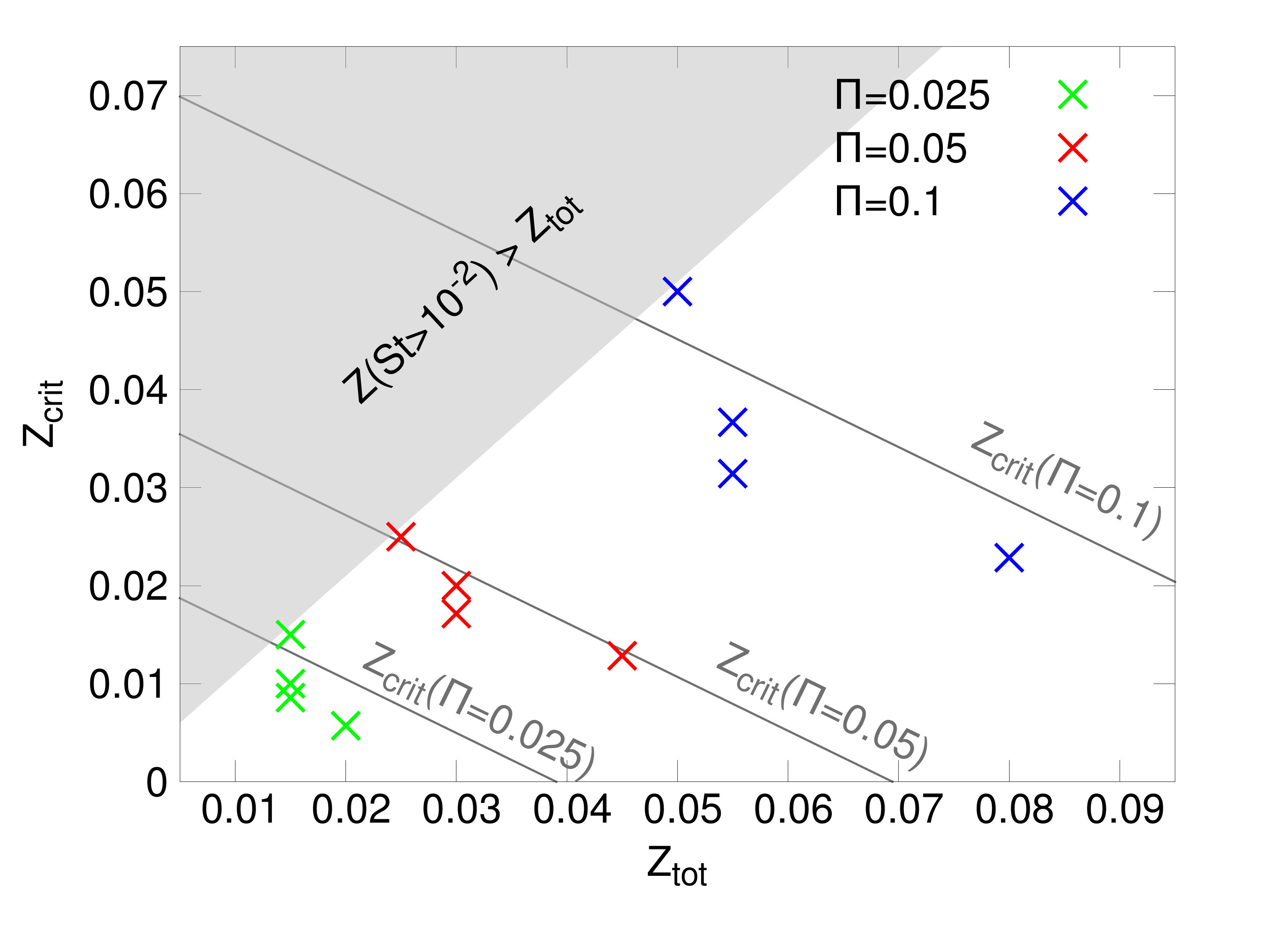}
      \caption{The critical metallicity of large aggregates necessary to trigger planetesimal formation as a function of total metallicity. The points present data obtained in direct numerical simulations by \citet{2010ApJ...722L.220B} for three different values of the $\Pi$ parameter. We note that \citet{2010ApJ...722L.220B} use the symbol $Z_{\rm crit}$ for the critical total metallicity they find for each run, whereas we use it for the metallicity of grains larger than ${\rm St_{crit}}=10^{-2}$. The lines show our fit to the data (Eq.~\ref{zcrit}). The shaded region is where the metallicity of large grains would be higher than the total one, which is not possible.
              }
         \label{fig:fit}
\end{figure}

The work of BS10 suggests that the value of $Z_{\rm crit}$ depends on the total metallicity $Z_{\rm tot}$ and the radial pressure gradient, which is parametrized with $\Pi$. We assume the dependence to be
\begin{equation}\label{zcrit}
Z_{\rm crit} =  {\rm a} \times Z_{\rm tot} + {\rm b} \times \Pi + {\rm c}.
\end{equation}
In order to determine values of the parameters ${\rm a}$, ${\rm b}$, and ${\rm c}$, we analyze the results presented by \citet{2010ApJ...722L.220B}. For each set of runs sharing the same size distribution and pressure support, we check what is the minimum total metallicity that allows planetesimal formation. Then, we calculate the metallicity contributed by the large grains in these runs ($Z_{\rm crit}$) and compare it to the total metallicity ($Z_{\rm tot}$). By numerical fitting to the data, we find ${\rm a}=-0.88$, ${\rm b}=0.912714$, and ${\rm c}=0.004125$. We present the fit and data extracted from \citet{2010ApJ...722L.220B} in Fig.~\ref{fig:fit}. With these values, we find that the absolute minimum metallicity that triggers strong clumping, for the standard value of $\Pi=0.05$, is $Z_{\rm crit} = 0.026$ (assuming that all the particles are large), which is significantly higher than the standard solar value of 0.01. This result is consistent with the findings of \citet{2009ApJ...704L..75J}.

Our planetesimal formation algorithm works as follows. At every advection time step of the Monte Carlo coagulation calculation, we check the mass distribution of grains produced by the interplay of settling and coagulation, and calculate the metallicity of large grains $Z({\rm St}>10^{-2})$. If the $Z({\rm St}>10^{-2})$ is higher than the threshold value of $Z_{\rm crit}$, planetesimal formation may happen. We also check whether the dust-to-gas ratio in the midplane exceeds unity, as this is a general condition for the streaming instability to be triggered. If both the conditions are fulfilled, we remove some amount of our largest aggregates, which corresponds to an assumed mass of clump $M_{\rm clump}$, from the dust coagulation code and refer to them as planetesimals. The evolution of the newly formed planetesimals is not included in the current version of the code.

The BS10 models did not include self-gravity, thus the formation of planetesimals was not followed. In our models, the quantity of dust grains removed in one planetesimal-forming event is estimated with the mass of a collapsing clump
\begin{equation}\label{mclump}
M_{\rm clump}=\rho_{\rm d}^{0} \times H_{\rm{d}}^3,
\end{equation}
where $\rho_{\rm d}^{0}$ is the dust density in the midplane and $H_{\rm{d}}$ is the vertical scale-height of the dust. For MMSN at 5~AU, we get $M_{\rm clump}\approx10^{22}$~g, which corresponds to 100 km-sized planetesimals with the internal density of 1~g~cm$^{-3}$, consistent with constraints from the asteroid belt \citep{2009Icar..204..558M}. Equation~(\ref{mclump}) presents a very crude order-of-magnitude estimate based on the typical height of the dust layer, which results from the interplay of settling and the turbulence driven by the streaming instability (Eqs.~\ref{hdust} and \ref{alpha}). In reality, the streaming instability forms elongated filaments that fragment to form planetesimals. Recently, \citet{2014ApJ...792...86Y} estimated the width of the filaments from direct numerical simulations to be on the order of $10^{-2}\times H_{\rm g}$, which is consistent with our $H_{\rm d}$ estimate (for ${\rm St}=10^{-2}$ and $\alpha=10^{-6}$). In their simulations, the filaments include dust of density $\sim 10^2\times \rho_{\rm g}^{0}$, where $\rho_{\rm g}^{0}$ is the gas density in the midplane. As in our models $\rho_{\rm d}^{0}\gtrsim\rho_{\rm g}^{0}$, the $M_{\rm clump}$ estimated by \citet{2014ApJ...792...86Y} would be by 2 orders of magnitude higher than calculated from Eq.~\ref{mclump}. Nevertheless, we find that the final outcome of our models does not depend on the exact value of the $M_{\rm clump}$ as long as there are no resolution problems coming into play.

To avoid the resolution difficulties, we take care that the mass represented by one swarm $M_{\rm swarm}$ (Eq.~\ref{mswarm}) is lower than the mass of the collapsing clump $M_{\rm clump}$, meaning that we remove more than one representative particle to account for collapse of one clump. Thanks to this, the amount of dust removed in one step is not too high. At the same time, we need the $M_{\rm swarm}$ to be higher than the maximum mass of aggregate that can be produced with coagulation, which can be estimated thanks to Eq.~(\ref{maxst}). The representative particles approach only works when the physical particles are less massive than the mass of one swarm, because of the assumption that one swarm represents multiple physical particles.

\section{Preliminary estimates}\label{sub:pre}

We investigate whether the dust coagulation can produce aggregates that are large enough to trigger strong clumping in the streaming instability, which can then lead to planetesimal formation, and model the planetesimal formation when it is possible. We present results of our numerical simulations in Sect.~\ref{sub:results}, but first we motivate our choice of parameter space with simple estimates.

The size of dust grains that can be obtained by coagulation is limited by numerous effects, such as collisional physics and radial drift. On the other hand, existence of the large grains is crucial for the streaming instability and subsequent planetesimal formation.

The maximum size of aggregates depends mainly on the critical velocity above which particles do not stick. 
Fragmentation velocities of silicate aggregates are measured in laboratory experiments to be in the range of a few tens of cm~s$^{-1}$ to a few m~s$^{-1}$, while bouncing collisions already happen at velocities of a few cm~s$^{-1}$ \citep{2010A&A...513A..56G,2013A&A...551A..65S,2014ApJ...783..111K}. The collision outcome is known to depend strongly on the porosity, and porous grains may grow even at velocities of 30 m s$^{-1}$ \citep{2011ApJ...737...36W,2013A&A...559A..62W, 2013MNRAS.435.2371M}. However, numerical models including the porosity have shown that the porous silicate grains will be collisionally compacted and the growth will be halted by bouncing \citep{2010A&A...513A..57Z}.
The dust grains consisting of ices are considered to be significantly more sticky and resistant to compaction \citep{2012ApJ...752..106O, 2013A&A...554A...4K, 2014MNRAS.437..690A}. However, as the laboratory experiments involving ices are more challenging than for silicate grains, there is still no detailed collision model for such grains. In the molecular dynamics models, ice grains are found to be very porous and able to grow even at velocities of 50 m s$^{-1}$ \citep{2009ApJ...702.1490W}.

To account for the difference in growth efficiency of silicates and ices in our simple model, we assume a different critical velocity for growth inside and outside the snow line. For the silicate particles present inside the snow line (located at 3~AU in this model) we assume an impact velocity for bouncing$\slash$fragmentation of $v_{\rm in}= 10$ cm~s$^{-1}$ and for icy particles $v_{\rm out}= 10$ m s$^{-1}$. 

\begin{figure}
   \centering
   \includegraphics[width=\hsize]{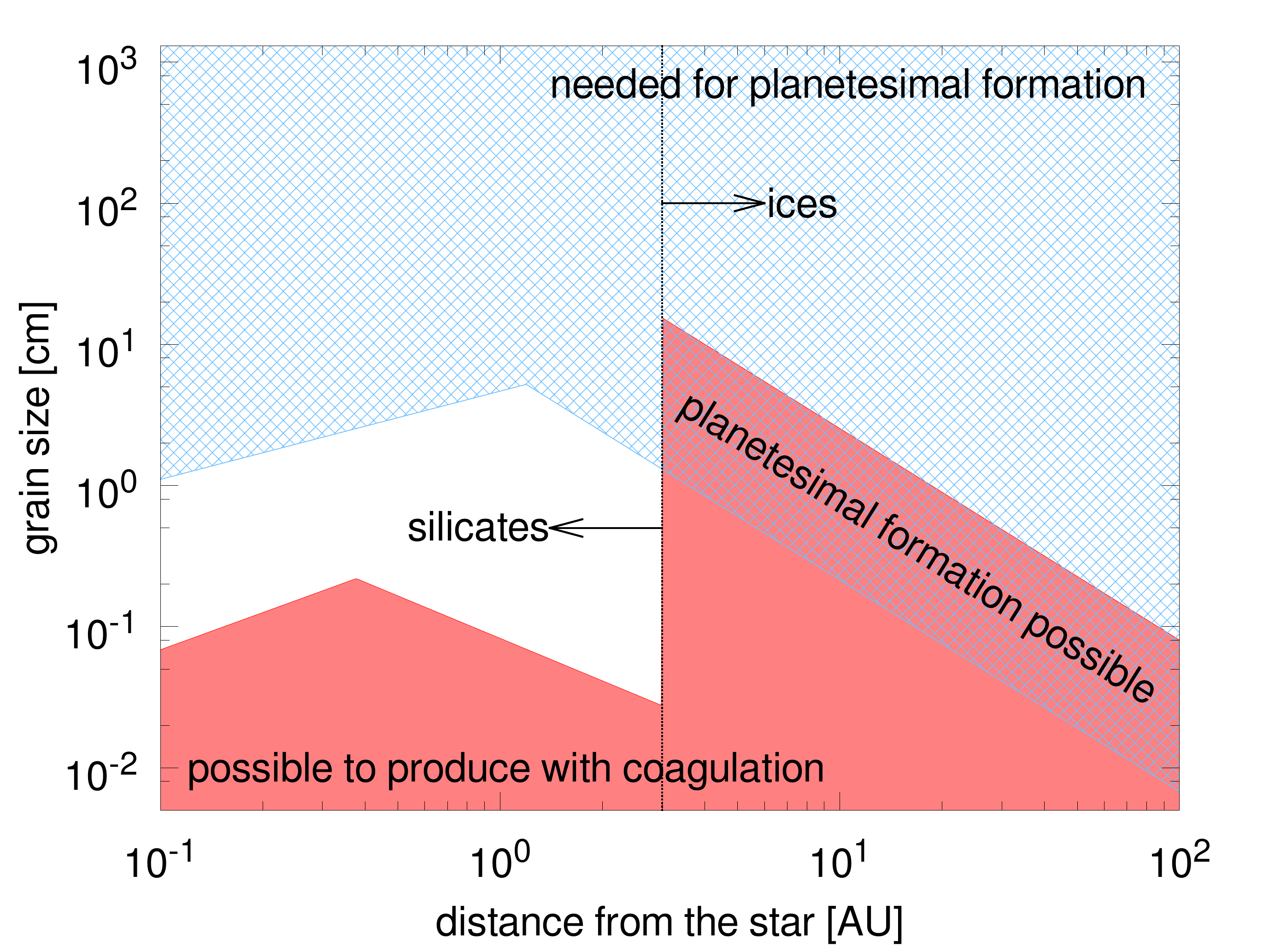}
      \caption{Comparison of maximum dust particle size produced by coagulation (red shaded region) and minimum size needed for planetesimal formation in a dead zone (blue crosshatched region). We assume that particle growth is limited by the relative drift velocities, ignoring turbulence, which corresponds to the dead zone. We find that the maximum particle size exceeds the size corresponding to the ${\rm St}=10^{-2}$ only for the ices that can exist beyond the snow line, where the presence of ice makes the dust particles more sticky.
              }
         \label{fig:sizes}
\end{figure}

As we place our models in a dead zone, we do not consider turbulence to be a source of impact velocities. The impact velocities are thus determined by relative drift, which can be parametrized by~$\eta$ (Eq.~\ref{eta}). Besides the bouncing and fragmentation, the maximum size of grains can also be restricted by removal of material with radial drift. However, we find that for the enhanced dust abundance, which is a prerequisite for an efficient streaming instability, the growth rate is enhanced as well, and the drift barrier does not influence the maximum size of grains, as this would only occur for particles larger than the size limited by fragmentation. We neglect the removal of material by the radial drift in this paper.

The maximum Stokes number of aggregates resulting from the collisions driven by relative drift can be estimated as \citep{2012A&A...539A.148B} 
\begin{equation}\label{maxst}
{\rm St}_{\rm max}\approx\frac{v_{\rm f}}{|\eta| v_{\rm K}}.
\end{equation}
We plot the size of aggregates corresponding to the ${\rm St}_{\rm max}$ and the critical value of ${\rm St_{crit}}=10^{-2}$ in the laminar MMSN disk in Fig.~\ref{fig:sizes}. In other words, Fig.~\ref{fig:sizes} shows where the dust growth can produce aggregates that are large enough to trigger the streaming instability.
We find that to obtain grains with ${\rm St}>10^{-2}$, the velocity $v_{\rm f}$ has to be typically higher than $\sim$1 m~s$^{-1}$.  
Thus, obtaining the particles of ${\rm St}>10^{-2}$ is very hard inside the snow line, where the growth of silicate particles is halted by bouncing. However, it should be relatively easy in regions where solid ice can exist. The snow line location is fixed at 3~AU in the simple model presented in this section. In a realistic disk, the snow line migrates with time \citep{2005ApJ...620..994D, 2011Icar..212..416M, 2012MNRAS.425L...6M, 2014arXiv1408.1016B}. If the snow line moves inward, the region where planetesimals can form extends.

\section{Results}\label{sub:results}

We use our dust evolution code together with the planetesimal formation prescription described in Sect.~\ref{sub:model} to model dust coagulation and planetesimal formation in a dead zone of the MMSN disk. Motivated by the estimates presented in the previous section, which show that the streaming instability can only form planetesimals outside the snow line, we locate our numerical models at 5~AU, where the cores of giant planets in the solar system were presumably formed. We assume that the dust aggregates have internal density of $\rho_{\rm p}=1$~g~cm$^{-3}$ and we treat them as compact spheres. The dust grains have an initial size of 1 $\mu$m and are distributed such that the dust-to-gas ratio is constant within the whole vertical range. We let the grains stick, fragment, settle down toward the midplane, and be stirred by turbulence triggered by the streaming instability when the dust-to-gas ratio in the midplane reaches unity. Fragmentation occurs for collisions with impact speeds higher than a critical value $v_{\rm f}$. All our runs have vertical resolution of 100 grid cells and we place 400 representative particles in each cell. For this level of resolution, we only find minor differences between runs started with identical parameters but different random seeds, as is usually practiced for Monte Carlo methods. 

\subsection{Fiducial run}\label{sub:tr}

For our fiducial run we choose $\Pi=0.05$, corresponding to a pressure gradient slightly reduced with respect to the nominal MMSN model, where $\Pi\approx0.08$ at 5~AU. However, this value matches a more realistic disk model presented by \citet{2010AREPS..38..493C}. In general, the value of $\Pi$ increases with the radial distance from the star. In the MMSN model $\Pi \approx 0.055\times(r/{\rm AU})^{1/4}$, whereas in the \citet{2010AREPS..38..493C} model $\Pi \approx 0.036\times(r/{\rm AU})^{2/7}$. We start the run with the vertically integrated dust-to-gas ratio of $Z=0.05$, a~factor of five higher than the usual solar metallicity. For the impact velocity above which the particles fragment, we take $v_{
\rm f}=1000$~cm~s$^{-1}$. A setup of this kind could correspond to a pressure trap induced by a long-lived zonal flow \citep{2013ApJ...763..117D}. 

\begin{figure}
   \centering
   \includegraphics[width=\hsize]{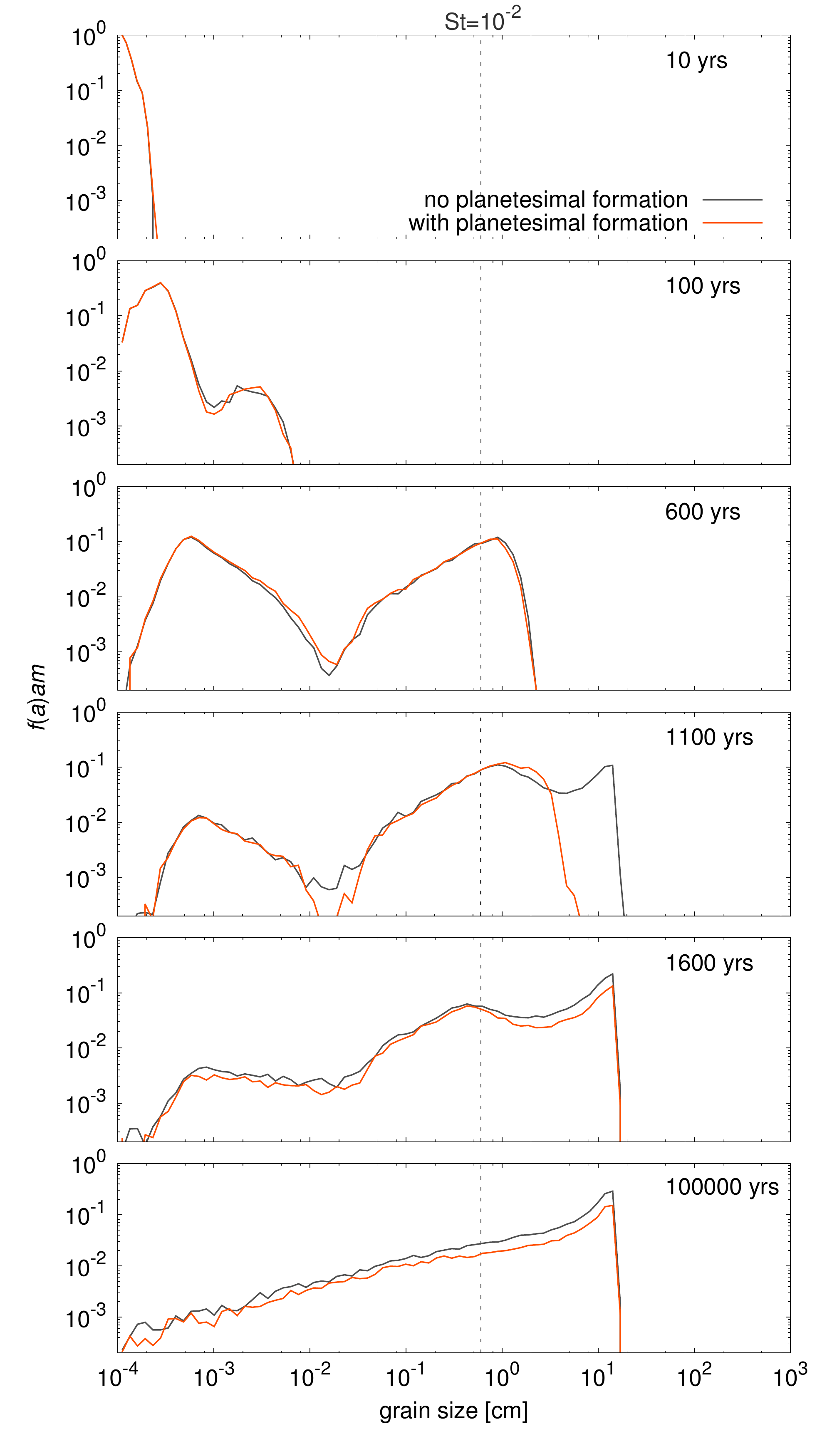}
      \caption{Time evolution of the dust size distribution in our fiducial run with and without the planetesimal formation by the streaming instability enabled. The vertical dashed line indicates grain size corresponding to a~Stokes number of 10$^{-2}$. Larger grains can form planetesimals and be removed from the simulation. 
              }
         \label{fig:massevo}
\end{figure}

Figure \ref{fig:massevo} shows the evolution of grain size distribution in the fiducial run. For comparison, we also show the evolution of the same run but without the planetesimal formation algorithm enabled. A~typical sedimentation-driven coagulation scenario happens: the equal-sized particles initially grow slowly thanks to the Brownian motion. Then, the particles in the upper layers start to grow much faster than those in the midplane, because the settling velocities increase with height (Eq.~\ref{vsett}). The largest aggregates sweep-up smaller particles while they settle down and thus further increase their settling velocity, resulting in formation of a bimodal size distribution at $\sim$600 yrs. Then, the particles encounter the fragmentation barrier and a coagulation-fragmentation equilibrium develops, leading to a power-law-like size distribution. The slope of the distribution depends on mass distribution of fragments implemented. We implement the fragment distribution $n(m)\ dm\propto~m^{-11/6}\ dm$, corresponding to the MRN distribution \citep{1977ApJ...217..425M}.
We find that the coagulation-fragmentation equilibrium in the dead zone, where collisions are mainly driven by the systematic drift, leads to the size distribution 
\begin{equation}\label{sizedistr}
n(a)\cdot a\cdot m\ d{\log}a \propto a^{1/2}\ d{\log}a,
\end{equation}
visible in the bottom panel of Fig.~\ref{fig:massevo}.
 
The evolution proceeds identically in both cases: with and without planetesimal formation, until $\sim$1000~yrs. After this time, both the conditions described in Sect.~\ref{sub:si} are fulfilled and some particles are removed to account for the planetesimal formation. Removing the large grains slows down the collisional evolution, which can be seen in the fourth panel of Fig.~\ref{fig:massevo}, because the missing large grains possess very high interaction rates and would normally participate in many collisions. However, this effect is quickly smeared out by the development of the coagulation-fragmentation equilibrium. The final difference between the size distributions is not very pronounced.

\begin{figure}
   \centering
   \includegraphics[width=0.9\hsize]{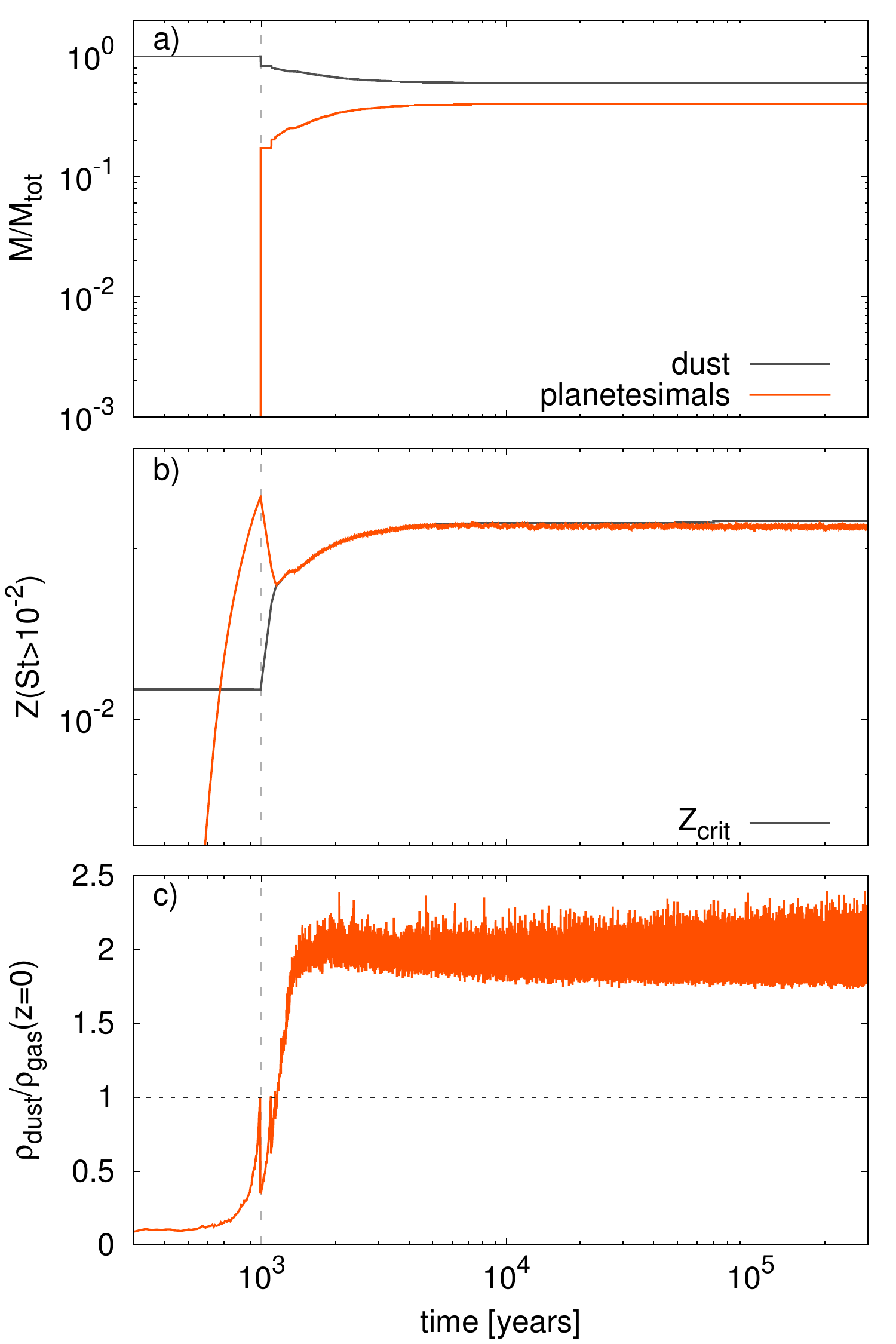}
      \caption{Time evolution of our fiducial run: a)~total planetesimals and dust mass, b)~ratio of the surface density of particles larger than ${\rm St}=10^{-2}$ to the gas density compared to the threshold value $Z_{\rm crit}$, and c)~dust-to-gas ratio in the midplane compared to the threshold value of unity. The dashed vertical line corresponds to the time when the first planetesimals are formed. It can be seen that although the metallicity of large grains was already higher than $Z_{\rm crit}$ at $t\sim$700~years (b), the planetesimal formation only starts at $t\sim$990~years (a) because the large grains have not settled to the midplane yet (c).
              }
         \label{fig:P5Z5V1000}
\end{figure}

Figure \ref{fig:P5Z5V1000} presents an extended overview of the fiducial run time evolution. Panel a) presents how the dust and planetesimals abundance changes with time. Planetesimal formation starts at $t\sim$990~years and lasts $\sim6\times10^4$ years. Planetesimal formation efficiency decreases over time. Panels b) and c) correspond to the two conditions that have to be fulfilled simultaneously to allow planetesimal formation: high metallicity in the form of large grains ($Z({\rm St}>10^{-2})>Z_{\rm crit}$) and the midplane dust-to-gas ratio higher than unity. In the presented case, the first condition is already fulfilled at $t\sim$700~years, but the settling takes an additional $\sim300$ years which delays the onset of planetesimal formation. The large quantity of large grains already present results in a planetesimal formation outburst: around 40\% of the final number of planetesimals is formed in the first planetesimal-forming step. After this, the turbulence generated by the streaming instability stirs the midplane layer below the threshold dust density again. Interplay of growth and settling leads to refilling the reservoir of large grains. Each time a clump is removed, the total metallicity $Z_{\rm tot}$ decreases and the threshold metallicity $Z_{\rm crit}$ increases (Eq.~\ref{zcrit}). After $\sim6\times10^4$ years, $Z_{\rm crit}$ is so high that the coagulation cannot produce enough large grains and planetesimal formation is no longer possible.

The fact that the sedimentation takes longer than growth up to the required sizes is not a general rule. In some of the other runs, presented in the following section, we observe the opposite relation, i.e., that the growth takes longer than settling. This occurs because both timescales are comparable in the dead zone case. Thanks to this, the sedimentation-driven coagulation may happen, where the growth and settling timescales for the rain-out particles are equal.
   
It is worth noting that the modeling of sedimentation-driven coagulation requires the vertical structure to be included. The sedimentation-driven coagulation determines the initial timescale of growth. The process we describe cannot be modeled in detail using vertically integrated algorithms, as these assume that equilibrium between vertical settling and mixing is reached on a timescale shorter than the growth timescale. We find that planetesimal formation starts before this equilibrium is reached.

\subsection{Parameter study}

\begin{figure*}
   \centering
   \includegraphics[width=\hsize]{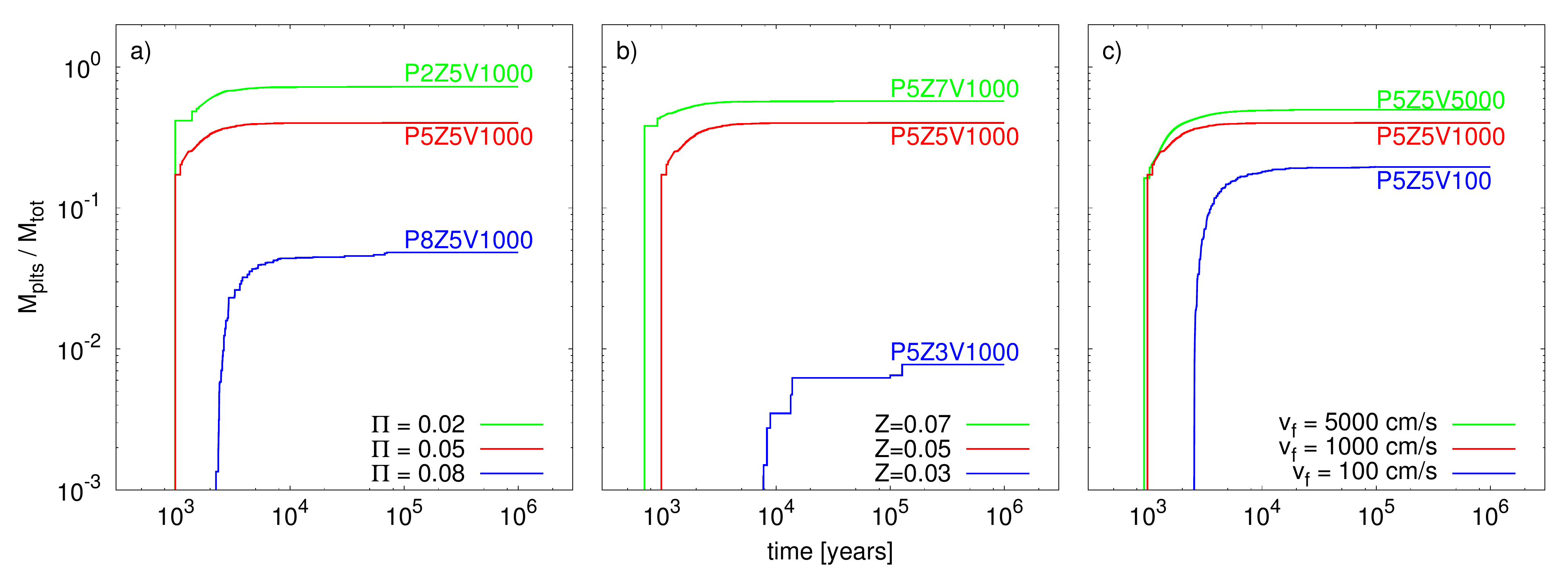}
   \caption{Planetesimal formation in our runs depends on parameters: a) radial pressure gradient parametrized by $\Pi$ (Eq.~\ref{pi}), b) total metallicity $Z_{\rm tot}$, c) fragmentation velocity $v_{\rm f}$. The red line in each panel corresponds to our fiducial run P5Z5V1000.}
   \label{fig:paramsstudy}
   \end{figure*}

In order to check how our results depend on the choice of parameters, we vary the fragmentation velocity $v_{\rm f}$, the pressure gradient $\Pi$, and the total vertically integrated metallicity $Z_{\rm tot}$. We perform 24 runs in total. Table~\ref{table:all} gives an overview of the parameter values used in our simulations and their results in terms of how much dust is turned into planetesimals and how long we have to wait for the planetesimal formation to be triggered. Our fiducial run is referred to as P5Z5V1000 in this table. Different panels of Fig.~\ref{fig:paramsstudy} show how the planetesimal formation changes when one of the parameters is varied from its fiducial value. We discuss the parameter dependencies further in this section. 

\begin{table}
\caption{Overview of parameters used and results obtained in different runs}
\centering                         
\begin{tabular}{l c c c c c}   
\hline\hline                
Run name & $\Pi$\tablefootmark{a} & $Z_{\rm tot}$\tablefootmark{b} & $v_{
\rm f}$\tablefootmark{c} & $M_{\rm plts}${\slash}$M_{\rm tot}$\tablefootmark{d} & $t_{\rm plts}$\tablefootmark{e} \\   
\hline
  P2Z7V5000  &      &      & 5000  & 83\% & 60      \\
  P2Z7V1000  & 0.02 & 0.07 & 1000  & 80\% & 64      \\
  P2Z7V100   &      &      & 100   & 78\% & 174     \\
\hline
  P2Z5V5000  &      &      & 5000  & 77\% & 85      \\
  P2Z5V1000  & 0.02 & 0.05 & 1000  & 73\% & 88      \\
  P2Z5V100   &      &      & 100   & 68\% & 212     \\
\hline
  P2Z3V5000  &      &      & 5000  & 61\% & 147     \\
  P2Z3V1000  & 0.02 & 0.03 & 1000  & 54\% & 151     \\
  P2Z3V100   &      &      & 100   & 47\% & 275     \\
\hline
  P5Z7V5000  &      &      & 5000  & 64\% & 59      \\
  P5Z7V1000  & 0.05 & 0.07 & 1000  & 57\% & 63      \\
  P5Z7V100   &      &      & 100   & 43\% & 175     \\
\hline
  P5Z5V5000  &      &      & 5000  & 50\% & 83      \\  
  P5Z5V1000  & 0.05 & 0.05 & 1000  & 40\% & 88      \\  
  P5Z5V100   &      &      & 100   & 20\% & 227     \\
\hline
  P5Z3V5000  &      &      & 5000  & 16\% & 233     \\
  P5Z3V1000  & 0.05 & 0.03 & 1000  & 0.8\%& 693     \\
  P5Z3V100   &      &      & 100   & 0\%  & -       \\
\hline
  P8Z7V5000  &      &      & 5000  & 35\% & 56      \\
  P8Z7V1000  & 0.08 & 0.07 & 1000  & 32\% & 61      \\
  P8Z7V100   &      &      & 100   & 0\%  & -       \\
\hline
  P8Z5V5000  &      &      & 5000  & 9\%  & 125     \\
  P8Z5V1000  & 0.08 & 0.05 & 1000  & 5\%  & 201     \\
  P8Z5V100   &      &      & 100   & 0\%  & -       \\     
\hline\hline
\end{tabular}
\label{table:all}          
\tablefoot{Columns (a)-(c): parameters used in each run, columns (d)-(e): results obtained; \tablefoottext{a}{Radial pressure gradient measure (Eq.~\ref{pi})} \tablefoottext{b}{Total vertically integrated dust-to-gas ratio} \tablefoottext{c}{Fragmentation velocity in cm~s$^{-1}$} \tablefoottext{d}{Total final mass of planetesimals produced with respect to the initial total mass of dust}
\tablefoottext{e}{Time when the planetesimal formation started in local orbital timescales}}
\end{table}

\subsubsection{Dependence on the pressure gradient}
Both \citet{2007Natur.448.1022J} and \citet{2010ApJ...722L.220B} observed that with a lower pressure gradient it is easier to trigger the planetesimal formation. This effect may seem
counterintuitive, as the streaming instability is driven by the existence of the pressure gradient and the sub-Keplerian rotation of gas. However, the radial pressure gradient provides free energy that triggers the turbulence; therefore, a lower pressure gradient weakens the turbulence and promotes particle settling to higher densities (see Eq.~\ref{alpha}). In our model, the radial pressure gradient is parametrized by $\Pi$ (Eq.~\ref{pi}). Based on the BS10 results, the critical metallicity for planetesimal formation, $Z_{\rm crit}$, is dependent on $\Pi$ such that a lower pressure gradient promotes planetesimal formation (Eq.~\ref{zcrit}). What is more, with a lower $\Pi$, the collision speeds between particles are also lower and thus the maximum particle size and the corresponding Stokes number (${\rm St_{max}}$) are higher. This is because, as seen in Eq.~(\ref{maxst}), the maximum Stokes number we can obtain is inversely proportional to the maximum radial drift velocity, which is proportional to the radial pressure gradient. With the two effects combined, the effect of the radial pressure gradient is very strong. As can be seen in panel a) of Fig.~\ref{fig:paramsstudy}, lowering the pressure gradient by 60\% results in a greater than 80\% increase in the number of planetesimals that can be formed. On the other hand, by increasing the gradient by 60\%, we get a dramatic decrease in the number of planetesimals produced: from 40\% to only 5\% of the initial mass of dust.

\subsubsection{Dependence on the metallicity}
Increasing the total amount of dust available has a positive effect on the number of planetesimals that are formed. First of all, higher total metallicity directly translates into a larger quantity of large grains ($Z({\rm St}>10^{-2})$) that can participate in the strong clumping that can lead to planetesimal formation. As seen in Eq.~(\ref{zcrit}), higher total dust abundance lowers the critical vertically integrated dust-to-gas ratio $Z_{\rm crit}$ because the higher dust abundance reduces the turbulence triggered by the instability (Eq.~\ref{alpha}) and thus makes the clumping easier. The higher dust-to-gas ratio in the midplane also leads to a faster rotation of the gas and reduction of the difference between the gas and Keplerian rotation \citep{2010ApJ...722.1437B}. This has a~similar effect to reducing of the radial pressure gradient, which facilities the planetesimal formation, as discussed above. 
Finally, the higher metallicity speeds up the growth and the planetesimal formation is able to start earlier, as the growth timescale scales inversely with the vertically integrated dust-to-gas ratio \citep{2012A&A...539A.148B},
\begin{equation}
\tau_{\rm growth}\propto \left(Z_{\rm tot}v_{\rm K}\right)^{-1},
\end{equation}
where $v_{\rm K}$ is the Keplerian velocity. We observe that with metallicity $Z_{\rm tot}>0.03$, the growth proceeds so quickly that it is the settling timescale which determines the onset of planetesimal formation. This effect takes place in our fiducial run P5Z5V1000. With these combined positive effects, metallicity strongly influences the number of planetesimals formed. However, the impact of metallicity is nonlinear: increasing it by 40\% results in $>$40\% increase in the number of planetesimals formed, but decreasing it by 40\% with respect to the fiducial value results in almost no planetesimals being formed (panel b of Fig.~\ref{fig:paramsstudy}). The strong dependence on the vertically integrated dust-to-gas ratio is consistent with the results of direct numerical simulations \citep{2009ApJ...704L..75J,2010ApJ...722.1437B}.

\subsubsection{Dependence on the fragmentation velocity}
Dependence on the fragmentation velocity is the most straightforward. The value of $v_{\rm f}$ only influences the maximum size of particles we are able to obtain by coagulation. The maximum Stokes number ${\rm St}_{\rm max}$ increases proportionally to the fragmentation velocity, as can be seen in Eq.~(\ref{maxst}). 
The higher the fragmentation velocity, the more grains will have a~Stokes number above the critical value (${\rm St_{crit}}>10^{-2}$), and it is easier to reach the condition of the high metallicity of large grains ($Z({\rm St}>10^{-2})>Z_{\rm crit}$). We present the dependence of our results on the fragmentation velocity in panel c) of Fig.~\ref{fig:paramsstudy}. For the chosen values, the $v_{\rm f}$ does not influence the outcome as strongly as the other two parameters. However, as we show in Sect.~\ref{sub:pre}, with even lower fragmentation velocities we would not be able to produce any grains larger than the critical size. 
As most of the mass is found in large grains, which is connected to the fragmentation law we chose, we conclude that the fragmentation velocity (once it has permited growth of grains with a~Stokes number ${\rm St}>10^{-2}$) does not have a~very strong influence on the number of planetesimals that can be formed.

\subsection{Analytical model}\label{sub:expl}

We presented results of our numerical simulations following interplay between coagulation and planetesimal formation under different conditions in the disk. In this section, we construct a relatively simple formula that estimates the results and compute the efficiency of planetesimal formation in terms of the pressure gradient, initial vertically integrated dust-to-gas ratio, and fragmentation velocity.

As we describe in the previous sections, the most important property that determines whether planetesimal formation is possible is the quantity of dust grains with Stokes number higher than the critical value ${\rm St_{crit}}=10^{-2}$. If the Stokes number of the largest grains that can be produced by coagulation is ${\rm St_{max}>St_{crit}}$, the quantity of grains of interesting sizes can be estimated taking into account the dust mass distribution. The mass distribution we obtain when the coagulation and fragmentation are in equilibrium is described with Eq.~(\ref{sizedistr}). We emphasize that this particular slope is a result of the chosen distribution of fragments, which was described in Sect.~\ref{sub:tr}.

Taking into account that $m\propto{\rm St}^3$ (in the Epstein drag regime, which applies to the small grains that we obtain in our simulations) and $d{\log}m=dm/m$, we can rewrite the size distribution (Eq.~\ref{sizedistr}) in terms of the Stokes number 
\begin{equation}\label{stokesdistr}
n({\rm St})\cdot m\ d{\rm St} \propto {\rm St}^{-1/2}\ d{\rm St}.
\end{equation}
The relative amount of dust above a critical Stokes number ${\rm St}_{\rm crit}$ is 
\begin{equation}\label{ratio}
\frac{Z({\rm St}>10^{-2})}{Z_{\rm tot}} = \frac{\int_{{\rm St}_{\rm crit}}^{{\rm St}_{\rm max}} n({\rm St})m\ d{\rm St}}{\int_{{\rm St}_{\rm min}}^{{\rm St}_{\rm max}} n({\rm St})m\ d{\rm St}} = \frac{\sqrt{\rm St_{max}}-\sqrt{\rm St_{crit}}}{\sqrt{\rm St_{max}}-\sqrt{\rm St_{min}}}.
\end{equation}
The minimum Stokes number ${\rm St_{min}}$ is in principle determined by the size of monomers, and for $\mu$m-sized grains ${\rm St_{min}}\approx10^{-6}$. The maximum Stokes number ${\rm St_{max}}$ is determined by the fragmentation velocity $v_{\rm f}$ and the impact speeds. In the dead zone case, collisions are mainly driven by radial, azimuthal and vertical drift and the maximum size of grains is estimated accurately by Eq.~(\ref{maxst}). We assume the critical Stokes number of ${\rm St}_{\rm crit}=10^{-2}$, as discussed in Sect.~\ref{sub:si}.

The planetesimal formation happens as long as the abundance of large grains is higher than critical ($Z({\rm St}>10^{-2}) > Z_{\rm crit}$). The threshold abundance $Z_{\rm crit}$ is dependent on the pressure gradient $\Pi$ and the total metallicity $Z_{\rm tot}$, as described with Eq.~(\ref{zcrit}). When planetesimals are produced, the total metallicity decreases and thus the critical metallicity increases. The planetesimal formation stops when the quantity of large grains, corresponding to $Z({\rm St}>10^{-2})$ cannot reach the threshold value $Z_{\rm crit}$ anymore. Taking into account the ratio of the mass included in large grains to the total mass resulting from the equilibrium mass distribution (Eq.~\ref{ratio}), we can derive a critical total vertically integrated dust-to-gas ratio $Z_{\rm tot,crit}$ below which planetesimal formation is no longer possible.
Using Eqs.~(\ref{zcrit}) and (\ref{ratio}), we obtain
\begin{equation}
Z_{\rm tot,crit} = \left({{\rm b}\Pi + {\rm c}}\right)\times\left({{ \dfrac{\sqrt{\rm St_{max}}-\sqrt{\rm St_{crit}}}{\sqrt{\rm St_{max}}-\sqrt{\rm St_{min}}} }-{\rm a}}\right)^{-1}.
\end{equation}
The values of ${\rm a}$, ${\rm b}$, and ${\rm c}$ are given in Sect.~\ref{sub:si}, and the values of the Stokes numbers are discussed under Eq.~(\ref{ratio}).

The end of the planetesimal formation phase happens when the total metallicity $Z_{\rm tot}$ has dropped below the critical total metallicity $Z_{\rm tot,crit}$ and the relative number of planetesimals produced $M_{\rm plts}/M_{\rm tot}$ can be estimated as
\begin{equation}\label{mpltslast}
\frac{M_{\rm plts}}{M_{\rm tot}} = \frac{Z_{\rm tot,0}-Z_{\rm tot,crit}}{Z_{\rm tot,0}},
\end{equation}
where ${Z_{\rm tot,0}}$ is the initial total metallicity.

\begin{figure}
   \centering
   \includegraphics[width=0.95\hsize]{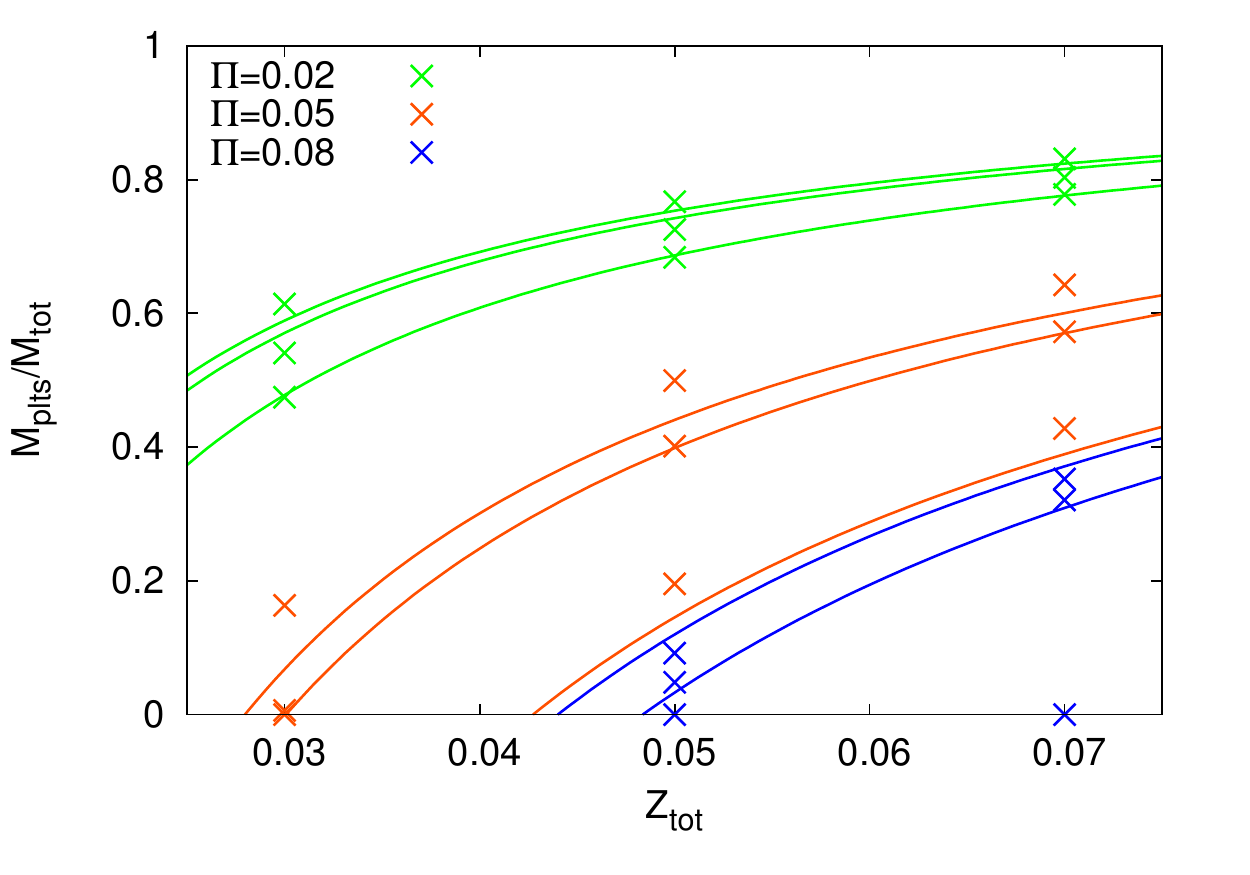}
      \caption{Comparison of the mass in planetesimals as a function of total metallicity obtained in different runs (points) and predicted by our analytical model (lines). The three different colors correspond to different values of the pressure gradient parameter $\Pi$. The three lines of the same colors correspond to different values of the fragmentation velocity used. For $\Pi=0.08$ and $v_{\rm f}=100$ cm~s$^{-1}$, there is no planetesimal formation predicted, thus the line is not visible.
              }
         \label{fig:toymodel}
   \end{figure}

Figure \ref{fig:toymodel} shows a comparison of the relative planetesimal mass produced, $M_{\rm plts}/M_{\rm tot}$, obtained with Eq.~(\ref{mpltslast}) and measured in our simulations (Table~\ref{table:all}). The efficiency of planetesimal formation given by Eq.~(\ref{mpltslast}) tends to slightly underestimate the results of our simulations. We associate this with the mass distribution peak around maximum grain size visible in the bottom panel of Fig.~\ref{fig:massevo}, which is not covered by the simple power law (Eq.~\ref{stokesdistr}) that we used to derive Eq.~(\ref{mpltslast}). This peak arises because of the fragmentation barrier: particles that grow mostly in roughly similar-sized collisions, suddenly lack slightly larger collision partners. In order to sustain a steady state, the particles pile up around the maximum size \citep{2011A&A...525A..11B}. 
   
\section{Discussion}\label{sub:discuss}

The work we have presented includes some inevitable assumptions. In this section, we discuss uncertainties contributed by these assumptions as well as possible ways to improve our work.

The collision model used in our work is highly simplified, restricting all the collision physics to one crucial parameter: the fragmentation velocity $v_{\rm f}$. What is more, we assume that all the dust grains are spherical and compact, with a constant internal density. It is known that the internal structure of grains is important to the coagulation \citep{2007A&A...461..215O,2009ApJ...707.1247O,2013MNRAS.435.2371M} and dynamics \citep{2012ApJ...752..106O,2014Icar..237...84H}. However, we lack reliable models of icy aggregate porosity, because laboratory experiments including the ices are particularly challenging. Our work may be improved in the future, as soon as laboratory data on collisions of icy aggregates are available.

Our streaming instability model relies on the work of BS10 and thus inherits all its uncertainties. The highest uncertainty is probably contributed by the fact that we focus on their 2-D simulations results, as due to computational expense reasons, the extended parameter study was performed in 2-D only. \citet{2010ApJ...722.1437B, 2010ApJ...722L.220B} reported that the conditions for the strong clumping are more stringent in 3-D than in 2-D, thus our model may be a bit too optimistic. On the other hand, the BS10 models do not include the particle's self-gravity, which might help to obtain strong clumping for lower metallicity or higher pressure gradient. This is because with the self-gravity a dust clump is able to collapse even if its initial density is lower than the Roche density \citep{2010ApJ...719.1021M}.

\citet{2010ApJ...722.1437B, 2010ApJ...722L.220B} reported that their runs saturate within $\sim$50-100 orbits. This is the time the dust needs to settle and the streaming instability needs to develop. The settling timescale is self-consistently included in our models, but we neglect the time the streaming instability needs to develop: we assume the streaming instability is able to produce planetesimals immediately after the dust has settled and a sufficient quantity of large grains is present. We motivate this assumption by the fact that the timescale of dust coagulation is typically longer than the timescale to develop the instability. What is more, the timescale for clump collapse measured in simulations that include self-gravity is very short, on the order of one orbit \citep{2011A&A...529A..62J}.

\citet{2010ApJ...722.1437B, 2010ApJ...722L.220B} assumed a flat mass distribution in logarithmic bins in the Stokes number, whereas in our models the mass distribution is a result of realistic coagulation-fragmentation cycle and roughly follows the power law described with Eq.~(\ref{sizedistr}). The model we built to overcome the problem of incompatible size distributions relies on splitting the dust mass into particles larger and smaller than a size corresponding to the ${\rm St_{crit}}=10^{-2}$. We assume that the large particles actively drive the instability and participate in clumps, and fit the dependence of critical abundance of the large aggregates on the total metallicity (Sect.~\ref{sub:si}, Fig.~\ref{fig:fit}). Both the choice of a constant value of ${\rm St_{crit}}$, and our fits should be rather treated as a first approximation. Given the limited data of the BS10 papers, this is the best we can do. This model may be improved as soon as more detailed parameter studies of the streaming instability become available. What is necessary is a more systematic parameter scan of the 3-D streaming instability models, ideally going hand-in-hand with results from coagulation models to ensure that realistic particle size distributions are used in these 3-D models.

The BS10 results are restricted to laminar disks, without the turbulence, unlike the work of \citet{2007Natur.448.1022J}. It is known that some source of viscosity is necessary as a mechanism of the angular momentum transport, which enables the disk accretion on observed timescales \citep{1973A&A....24..337S}. The magnetorotational instability (MRI) was considered as a mechanism to drive turbulence that is a source of such viscosity \citep{1991ApJ...376..214B}. Thus, neglecting the turbulence may seem a serious restriction. However, many models of protoplanetary disks suggest that large regions of the disks can be free from the MRI \citep{1996ApJ...457..355G,2007ApJ...659..729T,2013ApJ...765..114D}. In this picture, the midplane is quiescent, but the upper layers may still be active. The turbulence from the active layers may stir the midplane, which we neglect in this paper. Recently it was shown that the MRI is inefficient or even completely suppressed in the inner regions of disks when all the non-ideal magnetohydrodynamic effects, such as the ambipolar diffusion and the Hall effect, are taken into account \citep{2013ApJ...772...96B,0004-637X-791-2-137,2014A&A...566A..56L}. The typical extent of this dead zone is from 1 AU to 10 AU, which covers the planetesimal formation region we investigate. Thanks to the existence of a dead zone, not only are the planetesimals able to form, but they also avoid destructive collisions and grow by the runaway growth \citep{2011MNRAS.415.3291G, 2012MNRAS.422.1140G, 2013ApJ...771...44O}. The evolution of the planetesimals and their interaction with the remaining dust is not yet included in our model.

The extent of the dead zone is regulated by the quantity of small dust grains, whose large surface-to-mass ratio enables the sweep-up of free electrons from the gas and thus decreases the ionization \citep{2000ApJ...543..486S,2006A&A...445..205I}. In the case of efficient growth of dust grains, the dead zone vanishes \citep{2009ApJ...698.1122O,2012ApJ...753L...8O}. In our model, we neglect the dead zone evolution, but as we keep a significant number of small grains because of the fragmentation barrier, we consider this a safe approach.

We place our models at 5 AU, relying on our estimates, which show that it is not possible to grow sufficiently large grains out of silicate aggregates when the bouncing barrier is acting (Sect.~\ref{sub:pre}). It was shown that the bouncing barrier can be overcome and large grains can grow when the fragmentation with mass transfer effect is taken into account \citep{2012A&A...544L..16W,2013ApJ...764..146G,2014A&A...567A..38D}. However, the impact velocity distribution that is necessary to produce the seed gains in those papers is contributed by the MRI turbulence, which is not present in a dead zone. The MRI turbulence would make it harder to trigger the streaming instability, as it does not allow a very thin midplane layer to form. Thus, the sweep-up growth scenario may be operating and forming planetesimals preferentially in the active zone of protoplanetary disk, whereas the scenario we investigate in this paper works better in the dead zone.

Another way of forming sufficiently large aggregates inside the snow line is co-accretion of dust and chondrules, suggested by \citet{2008ApJ...679.1588O} and subsequently investigated in laboratory experiments by \citet{2012Icar..218..701B} and \citet{2012A&A...542A..80J}. In this scenario, chondrules acquire dusty rims that facilitate their sticking and allow the bouncing barrier to be overcome. As shown by \citet{2008ApJ...679.1588O}, this scenario requires moderately low turbulence, which would also be favorable for planetesimal formation via the streaming instability.

Our models are local and neglect the radial drift. The radial drift timescale for pebbles of ${\rm St}=10^{-2}$ at 5 AU is on the order of 10$^3$ orbits, which is one order of magnitude longer than the coagulation needs to trigger the streaming instability, thus we consider this a safe approach.

We assume that metallicity is enhanced over the standard solar value of 0.01. Such an enhancement could form in a radial drift dominated disk, where the pebbles necessary to trigger the streaming instability are drifting inward and may form pile-ups, as suggested by \citet{2002ApJ...580..494Y}, \citet{2004ApJ...601.1109Y}, \citet{2012A&A...537A..61L}, and \citet{2014MNRAS.437.3037L}. However, such an enhancement typically happens inside the snow line, where growing sufficiently large particles is suppressed by the non-stickiness of aggregates and the existing large aggregates may be destroyed by ice evaporation and high-speed collisions. A way to overcome this problem may be the reduction of impact speeds in the presence of strong dust-to-gas ratio enhancements \citep{1986Icar...67..375N,2010ApJ...722.1437B}, which is not yet included in our code. The reduction of impact speeds would, in general, help to form larger aggregates and thus to produce more planetesimals.

Another possibility that would justify the enhanced metallicity and the reduced pressure support at the same time is the existence of some kind of pressure bump. However, the formation process and lifetime of such pressure bumps remains uncertain. Pressure bumps caused by zonal flows have been observed in numerical simulations including the MRI turbulence \citep{2009ApJ...697.1269J, 2010A&A...515A..70D,2011ApJ...736...85U}, but their lifetimes are up to 50 orbits \citep{2013ApJ...763..117D}.  A pressure bump arising around the snow line was suggested by \citet{2007ApJ...664L..55K}, but it was recently found to require an unrealistically high viscosity transition \citep{2014arXiv1408.1016B}. What is more, complicated evaporation and condensation processes have to be taken into account to model the size evolution of dust \citep{2011ApJ...739...18K} when considering the region near the snow line. Thus, our model is not self consistent in the pressure bump and dust enhancement origin aspect:
we start our runs with metallicity already enhanced by a factor of a few with respect to a nominal solar value, as otherwise the planetesimal formation by streaming instability is not possible. We investigate the interplay between vertical settling, coagulation, and planetesimal formation, ignoring the radial drift of dust. Such an enhancement would in reality build up over a timescale determined by the radial drift, even in the presence of a pre-existing pressure bump. We plan to investigate the effects of radial drift on our results in a future work. 

The sizes of clumps and planetesimals created by the streaming instability are highly uncertain, as the hydrodynamic simulations have limited resolution and are typically not able to follow the clump collapse with realistic collisional behavior. As described in Sect.~\ref{sub:si}, we assume that the collapsing clumps have identical mass, which is estimated based on the height of the dust layer (Eq.~\ref{mclump}). This is only an order-of-magnitude estimate, which is not necessarily consistent with the recent results of \citet{2014ApJ...792...86Y}, as discussed in Sect.~\ref{sub:si}. However, we find that, although the details of evolution are dependent on the assumed clump mass, the final outcome in terms of total planetesimal mass produced is not. 

\section{Conclusions}\label{sub:last}

The streaming instability was proposed as an efficient way of overcoming the growth and drift barriers and forming planetesimals. However, strong clumping was proven to require large grains. In this paper, we investigated whether large grains can form in sufficient amount during coagulation under realistic conditions. We developed and implemented a simplified model for planetesimal formation in our dust coagulation code, as described in Sect.~\ref{sub:model}. Our work is a step toward a unified model for planetesimal formation because we join the dust coagulation modeling with planetesimal formation via streaming instability for the first time. 

We find that planetesimal formation by streaming instability is hindered for the silicate aggregates because the bouncing barrier prevents growth to the sizes (Stokes numbers) needed to trigger the streaming instability. It is possible to obtain the minimum size of particles, corresponding to the critical Stokes number (${\rm St_{crit}}=10^{-2}$), for the stickier, icy aggregates, which are present beyond the snow line. If some way can be found to overcome the bouncing barrier, we can also expect the streaming instability to operate inside the snow line. However, the strong clumping may only be triggered when the vertically integrated dust-to-gas ratio is enhanced by a factor of at least three with respect to the solar value of 0.01 and$\slash$or the radial pressure gradient is reduced with respect to the standard minimum mass solar nebula model. What is more, a dense midplane layer of solids have to be formed, which is only possible if the turbulence is relatively weak.

We modeled the interplay of dust coagulation and settling and planetesimal formation and performed a parameter study, varying the radial pressure gradient, metallicity and fragmentation velocity. We investigated how these values influence the amount of planetesimals formed. We proposed a simple explanation of the obtained results, by constructing an analytical expression for the maximum number of dust that can be turned into planetesimals (Sect.~\ref{sub:expl}). This model can be used in future projects, for example to constrain initial conditions for planetesimal evolution models or in planet population synthesis codes.

\begin{acknowledgements}
We thank Chris Ormel, Carsten Dominik and Alessandro Morbidelli for useful discussions and encouragement. We thank Xue-Ning Bai and the referee, Anders Johansen, for their comments that helped us to improve this paper. J.D. would like to acknowledge the use of the computing resources provided by bwGRiD (http:$\backslash\backslash$www.bw-grid.de), member of the German D-Grid initiative, funded by the Ministry for Education and Research (Bundesministerium f\"{u}r Bildung und Forschung) and the Ministry for Science, Research and Arts Baden-Wuerttemberg (Ministerium f\"{u}r Wissenschaft, Forschung und Kunst Baden-W\"{u}rttemberg)
\end{acknowledgements}
\bibliographystyle{aa}
\bibliography{coagulation-si.bib}

\end{document}